# Adapting approximate memory potentials for time-dependent density functional theory

Yair Kurzweil and Roi Baer[•]
*Department of Physical Chemistry and the Fritz Haber Center for Molecular Dynamics,
the Hebrew University of Jerusalem, Jerusalem 91904 Israel.*

Frequency dependent exchange correlation kernels for time-dependent density functional theory can be used to construct approximate exchange-correlation potentials. The resulting potentials are usually not translationally covariant nor do they obey the so-called zero-force condition. These two basic symmetry requirements are essential for using the potentials in actual applications (even within the linear response regime). We provide two pragmatic methods for imposing these conditions. As an example we take the Gross and Kohn (GK) frequency dependent XC functional (Phys. Rev.Lett. **55**, 2850 (1985)), correct it, and numerically test it on a sodium metal cluster. Violation of the basic symmetries causes instabilities or spurious low frequency modes.

## I. INTRODUCTION

Time-dependent density functional theory (TDDFT)[1] is gaining recognition as a robust, accurate gentle scaling approach to computing excitation energies molecular systems[2]. The idea in TDDFT is almost identical to that of the parent, namely density functional theory (DFT). We replace the physical electron system (the so-called "interacting system") by a system of non-interacting Fermions (the so-called "non-interacting system"). While the interacting particles are subject to an external force derived from the Coulomb potential of the nucleus and the time-dependent laser field, $v_{ext}(\mathbf{r},t)$, the non-interacting particles "feel" a force derived from an effective potential $v_s(\mathbf{r},t)$ (we follow here the common practice of using the index $s$ for quantities in the non-interacting system). This potential is a unique functional of the density and is usually written as a sum:

$$v_s(\mathbf{r},t) = v_{ext}(\mathbf{r},t) + v_H[n](\mathbf{r},t) + v_{XC}[n](\mathbf{r},t) \quad (1)$$

Where $v_H[n](\mathbf{r},t) = \int n(\mathbf{r}',t)/|\mathbf{r}-\mathbf{r}'| d^3r'$ is the Hartree potential. The last term is the exchange correlation (XC) potential, which is in general an unknown functional of the density. The basic approximation of TDDFT is to have $v_{XC}[n]$ equal to an approximate ground state functional. This approximation is valid in the slow weakly time-dependent regime and is known as the adiabatic functional approximation. The adiabatic XC potentials are quite successful for calculating excitation energies of molecules. It seems their performance is robust and they give relatively good results even in cases beyond their regime of validity. A prototype example is the adiabatic local density approximation (ALDA). There are many cases where ALDA yields good and sometimes excellent results for molecules, clusters, metals, crystals etc.[2] However, in many other cases, ALDA's functionals are unable to capture crucial parts of the dynamics of the given system. For instance, the line broadening of the strong collective excitations in metal clusters and crystals (and even in the ideal case of homogeneous electron gas (HEG)) are missed by ALDA. Such broadening is related to the retardation effects which are resulted from the electron-electron interaction. Going beyond ALDA, and constructing nonadiabatic XC functional is now about 20 years old important challenge in TDDFT.

Only few exact properties of the exact XC potential are known. Therefore, it is a nontrivial task to improve the potential beyond its adiabatic part. Gross and Kohn (GK),[3] were the first to suggest an XC potential which in the linear response limit has abides to some exact dynamical properties of the homogeneous electron gas (HEG). However, the direct application of the GK potential to systems other than the HEG leads to violation of some exactly known basic symmetry laws[4,5]. These laws are derived from the fundamental notions of classical mechanical description of space and time, since the era of Galileo and Netwon. A complete treatment of this problem leads one to the realm of time-dependent current-density functional theory (TDCDFT)[6-9]. However, the problems can also be treated within TDDFT[5,10,11].

In TDDFT, the basic symmetries reduce to two constraints, as discussed in Section II. This paper centers on the ways a given potential (such as the Gross-Kohn) can be implemented in a way that obeys the basic principles. Our previous work on this issue[10] relied on the use of an action functional which was Galilean invariant[12] and is now extended in two senses. First, we describe methods for adapting the XC potentials which do not depend on an action. Second, we compare the numerical performance of the GK functional with and without the imposing the basic symmetries. We find that lack of observance of the symmetries may cause the appearance of spurious modes or instabilities.

The two basic principles we address, which must be obeyed in TDDFT were first described by Vignale.[5] We consider a density distribution of electrons in two interacting system starting from its ground state. One principle results from Galilean covariance and considers two observers, moving one with respect to the other, with coordinate system $\mathbf{R}'$ and $\mathbf{R}$ respectively ($\mathbf{R} = \mathbf{R}' + \mathbf{x}(t)$). Both are watching the same electronic density, described as $n(\mathbf{R},t)$ by one observer and $n'(\mathbf{R}',t)$ by the second:

$$n'(\mathbf{R}',t) = n(\mathbf{R}' + \mathbf{x}(t),t). \qquad (2)$$

According to Newton, the force $-\nabla v_{ext}(\mathbf{R},t)$ measured by one observer is related to that measured by the second by an additive inertial force (assuming unit mass):

$$-\nabla v'_{ext}(\mathbf{R}',t) = -\nabla v_{ext}(\mathbf{R}' + \mathbf{x}(t),t) + \ddot{\mathbf{x}}(t). \qquad (3)$$

When the two observers consider the non-interacting system they find the same rule:

$$-\nabla v'_s(\mathbf{R}',t) = -\nabla v_s(\mathbf{R}' + \mathbf{x}(t),t) + \ddot{\mathbf{x}}(t). \qquad (4)$$

Combining Eqs. (3), (4) and (1) gives (up to a purely time-dependent phase) the "translationally covariance" (TC) condition[5]:

$$v_{XC}[n'](\mathbf{R}',t) = v_{XC}[n](\mathbf{R} + \mathbf{x}(t),t). \qquad (5)$$

A different kind of constraint arises from the fact that the electron density $n(\mathbf{R},t)$ is the same in both interacting and non-interacting systems. The acceleration of the electronic center of mass is therefore identical in both systems and so according to Ehrenfest's theorem[13] the force expectation value must too be equal:

$$-\int n(\mathbf{r},t)\nabla v_{ext}(\mathbf{r},t)d^3r = -\int n(\mathbf{r},t)\nabla v_s(\mathbf{r},t)d^3r. \qquad (6)$$

Using Eq. (1) and the easily shown fact that the Hartree potential does not contribute to the expectation value of the force we find the second, "no XC force condition"[5] or for brevity, "no force condition":

$$\mathbf{E}_{XC}(t) \equiv -\frac{1}{N_e}\int n(\mathbf{R},t)\nabla v_{XC}(\mathbf{R},t)d^3R = 0 \qquad (7)$$

Here we have defined the average $XC$ force per particle, which must be zero. One point of interest is the behavior of $\mathbf{E}_{XC}$ even in approximate cases when a very weak time-dependent perturbation is applied. In this case, we are usually interested only in the linear response. We assume that the ground-state XC potential already obeys the zero force condition. This is indeed the case in all popular potentials. In this case it is possible to show that in general $\mathbf{E}_{XC}$ is a first order quantity, since

$$\begin{aligned}\mathbf{E}_{XC}(t) = &-\frac{1}{N_e}\int \delta n(\mathbf{R},t)\nabla v_{XC}(\mathbf{R})dR \\ &+\frac{1}{N_e}\int \nabla n(\mathbf{R})\delta v_{XC}(\mathbf{R},t)dR\end{aligned} \qquad (8)$$

One obvious exception is the homogeneous gas case, where to leading order $\mathbf{E}_{XC}$ is already zero. Thus, there is no need to impose the zero force condition in this special case.

In TDCDFT we require that both net torque and net force vanish. In TDDFT, the XC potential must be TC and comply to the zero XC force condition in order to satisfy the harmonic potential theorem (HPT)[4] as detailed in appendix A.

XC potentials which often go beyond the adiabatic approximation are often written in the following way:

$$v_{XC}[n](\mathbf{r},t) = v_A[n](\mathbf{r},t) + v_M[n](\mathbf{r},t) \qquad (9)$$

Where $v_A$ is an approximate adiabatic potential (say, ALDA) and $v_M[n]$ includes memory (nonadiabatic) effects.

## II. IMPOSING TRANSLATIONAL COVARIANCE

In this section we first survey the center of mass method to impose translational covariance. This method was developed by Vignale[5]. We then implement this method within the linear response Kernel.

Given a functional $V_{XC}[n]$, how one can convert it into a TC functional, in the sense of Eq. (5). in the following way[5, 10, 12] by using the concept of a "proper density" (i.e. the density as seen by an observer in the center of mass system). A proper density is a density defined with respect to a body fixed frame. For example, the density defined with respect to the electronic center of mass:

$$\mathbf{D}[n](t) \equiv \frac{1}{N_e} \int n(\mathbf{R}',t) \mathbf{R}' d^3 R', \quad (10)$$

Where $N_e$ is the number of electrons. In this case we have the obvious relation: $\delta \mathbf{D}(t)/\delta n(\mathbf{R}',t') = \delta(t-t')\mathbf{R}'/N_e$. A proper density is a density defined with respect to $\mathbf{D}(t)$, i.e. with respect to a coordinate $\mathbf{r}$ in the CM system (CMS):

$$\mathbf{r} = \mathbf{R} - \mathbf{D}[n](t), \quad (11)$$

Hence, a 'proper density' $N(\mathbf{r},t)$, in the CMS is:

$$N(\mathbf{r},t) \equiv n(\mathbf{r} + \mathbf{D}[n](t),t). \quad (12)$$

The relation between $\mathbf{D}$ in the unprimed and primed systems is clearly given by:

$$\mathbf{D}[n'](t) = \mathbf{D}[n](t) - \mathbf{x}(t), \quad (13)$$

And so, using Eqs. (2), (12) and (13) we show that the coordinate $\mathbf{r}$ and the proper density are translationally invariant (i.e. both observers will report the same value).

$$N[n'](\mathbf{r},t) = N[n](\mathbf{r},t). \quad (14)$$

Given any $V_{XC}(\mathbf{R},t)$, we consider it as the potential in the proper system and thus in any other system we can define the potential:

$$v_{XC}[n](\mathbf{R},t) = V_{XC}[N[n]](\mathbf{R} - \mathbf{D}[n](t),t) \quad (15)$$

Based on Eq. (15), it is easily verified that this potential is TC (i.e. obeys Eq. (5)).

In Appendix A we give the expressions needed to correct any linear response kernel so that it is TC. Now we turn to discuss the zero force condition

## III. IMPOSING THE NO FORCE CONDITION

In the previous section, we discussed the issue of translational invariance and surveyed ways of constructing TC potentials. In this section we discuss the issue of zero XC force.

We find that imposing this condition can be done in two ways. The first, is by introducing a compensating homogeneous field. The second is by introducing a non-homogeneous field which is optimized in some sense.

## COMPENSATING HOMOGENEOUS FIELD METHOD

Given a any $\tilde{v}_{XC}[n](\mathbf{R},t)$, it is straightforward to correct it to satisfy the zero XC force condition. In our previous approach to the subject we used a TI action to derive a potential which is TC and obeys the no force condition. We found that besides the TC terms, a compensating TD homogeneous electric field appears[10]:

$$v_{XC}[n](\mathbf{R},t) = \tilde{v}_{XC}(\mathbf{R},t) + \tilde{\mathbf{E}}_{XC}(t) \cdot (\mathbf{R} - \mathbf{D}(t)) \quad (16)$$

Where $N_e = \int n(\mathbf{R},t) d^3 R$ is the number of electrons and $\tilde{\mathbf{E}}_{XC}$ is the XC force per particle of $\tilde{v}_{XC}$:

$$\tilde{\mathbf{E}}_{XC}(t) = -\frac{1}{N_e} \int \nabla \tilde{v}_{XC}(\mathbf{Y}) n(\mathbf{Y}) d^3 R \quad (17)$$

where we now introduce a new notation, of a space-time point $\mathbf{Y} \equiv (\mathbf{R},t)$, to facilitate the equations.

The XC kernel for the homogeneous field correction is determined by a linear response analysis of Eq. (16). We denote the sensitivity of the force to a density perturbation at $\mathbf{Y}' = (\mathbf{R}',t')$ by:

$$\begin{aligned}\mathbf{Q}(t;\mathbf{Y}') &\equiv \frac{\delta}{\delta n(\mathbf{R}',t')} \tilde{\mathbf{E}}_{XC}(t) \\ &= -\frac{1}{N_e} \nabla \tilde{v}_{XC}(\mathbf{R}',t') \delta(t-t') \\ &+ \frac{1}{N_e} \int \tilde{f}_{XC}(\mathbf{R},t;\mathbf{R}',t') \nabla n(\mathbf{R},t) d^3 R\end{aligned} \quad (18)$$

where $\tilde{f}_{XC}(\mathbf{Y},\mathbf{Y}') = \delta \tilde{v}(\mathbf{Y})/\delta n(\mathbf{Y}')$. Assuming that Thus, the XC kernel of Eq. (16), by the HF method is:

$$F_{HFXC}(\mathbf{Y};\mathbf{Y}') = \tilde{f}_{XC}(\mathbf{Y};\mathbf{Y}') + \mathbf{Q}(t;\mathbf{Y}') \cdot (\mathbf{R} - \mathbf{D}_0) \quad (19)$$

Where $\mathbf{D}_0$ is the electronic center of mass in the ground-state system.

## MINIMAL TAMPERING APPROACH

The fix for no-force condition appearing in Eq. (16) involves a linear potential felt all over space. This seems somewhat inconvenient and perhaps artificial. It seems that a local term is more physical and convenient. To obtain a local potential we take the following approach. Given an arbitrary $\tilde{v}_{XC}(\mathbf{R},t)$, we would like to find a potential obeying the zero force condition but one which is as similar to it as possible. We call the new XC potential the minimal tampering XC (MTXC) potential. We define, therefore, the following functional $A[v]$ to be minimized including a zero force constraint:

$$A[v] = \frac{1}{2}\int d^3R |v(\mathbf{Y}) - \tilde{v}_{XC}(\mathbf{Y})|^2 + \vec{\lambda}\cdot\int d^3R\, n(\mathbf{Y})\nabla v(\mathbf{Y}), \quad (20)$$

Where, $\mathbf{Y} \equiv (\mathbf{R},t)$, as before and $\vec{\lambda}$ is a 3-vector of Lagrange multipliers. The stationary point of $A[v]$, $\delta A[v]/\delta v(\mathbf{R},t) = 0$, yields the minimal tampering potential $V_{MTXC}(\mathbf{R},t)$:

$$V_{MTXC}[n](\mathbf{Y}) = \tilde{v}_{XC}[n](\mathbf{Y}) + \vec{\lambda}(t)\cdot\nabla n(\mathbf{Y}). \quad (21)$$

Multiplying both sides of (21) by $\nabla n(\mathbf{R},t)$, integrating over space and solving for $\vec{\lambda}$ gives:

$$\vec{\lambda}(t) = -\overleftrightarrow{M}^{-1}\tilde{\mathbf{E}}_{XC}, \quad (22)$$

Where the $3\times 3$ symmetric matrix positive definite $\overleftrightarrow{M}$ is defined by the density gradient, independent of $\tilde{v}_{XC}$:

$$M_{ij}(t) = \frac{1}{N_e}\int \partial_i n(\mathbf{R},t)\partial_j n(\mathbf{R},t)d^3R \quad (23)$$

It is straightforward to prove from Eqs. (21), (22) that $V_{MTXC}(\mathbf{R},t)$ is TC, provided $\tilde{v}_{XC}(\mathbf{R},t)$ is so.

One possible difficulty with this method is when the density is nearly homogeneous, so that $M$ is close to singular. As discussed above (after Eq. (8)), in a perfectly homogeneous gas case the first order correction is zero. So, we need only stabilize $\lambda$ so that in the homogeneous case it gives a zero result. Therefore a Tikhonov regularization is in place, replacing the inverse of $\overleftrightarrow{M}$ in Eq. (22) by a regularized inverse:

$$\overleftrightarrow{M}(t)^{-1} \to -\left(\overleftrightarrow{M}^2 + \alpha I\right)^{-1}\overleftrightarrow{M} \quad (24)$$

Where $\alpha$ is a small positive regularization parameter.

What is the XC-kernel corresponding to the minimal tampering approach in Eq. (21)? Note that for non-homogeneous electron distribution the quantity $\lambda$ is already a first order quantity so:

$$F_{MTXC}(\mathbf{Y};\mathbf{Y}') = \tilde{f}_{XC}(\mathbf{Y};\mathbf{Y}') - \overleftrightarrow{M}^{-1}\mathbf{Q}(t;\mathbf{Y}')\cdot\nabla n(\mathbf{R}) \quad (25)$$

## IV. NUMERICAL APPLICATIONS

By studying simple model cases of sodium ionic clusters we now demonstrate some of the application issues involved in satisfying the basic requirements discussed above. Let us take an XC functional composed of the ALDA functional (which obeys these requirements) and add to it a memory part, $v_{Mem}(\mathbf{R},t)$:

$$v_{XC}[n](\mathbf{R},t) = v_{ALDA}(n(\mathbf{R},t)) + v_{mem}[n](\mathbf{R},t) \quad (26)$$

Here, $v_{LDA}(n) = \frac{d}{dn}(\varepsilon_{LDA}(n)n)$ where $\varepsilon_{LDA}$ is the energy per electron of a homogeneous electron gas. In order to specify the memory term, we use the linear response kernel suggested by Gross and Kohn[3], $\operatorname{Im}\tilde{f}_{GK}(n,\omega) = a\omega/(1+b\omega^2)^{5/4}$ where $a$ and $b$ depend on the density $n$, as given in refs [3] and [14]. The real part of $\tilde{f}_{GK}$ is determined using the Kramers Kroning relations (see Appendix D of ref.[7] for the details). By definition, the linear response around the ground state of a homogeneous electron gas gives:

$$\delta v_{XC}(n,t) = \int_0^t f_{GK}(n,t-t')\delta n(t')dt \quad (27)$$

Where the real-time GK kernel, is defined by:

$$f_{GK}(n,t) = \frac{1}{2\pi}\int_{-\infty}^{\infty}\tilde{f}_{GK}(n,\omega)e^{-i\omega t}d\omega \qquad t > 0 \quad (28)$$

In order to get a potential which is valid beyond linear response, but compatible with the linear response of the homogeneous electron gas, we first construct the memory potential:

$$v_{MEM}[n](\mathbf{R},t) = \int_0^t F_{GK}(n(\mathbf{R},t),t-t')\dot{n}(\mathbf{R},t')dt' \quad (29)$$

Where $F_{GK}(n,t)$ is defined as:

$$F_{GK}(n,t) = \int_0^t f_{GK}(n,t')dt' - v'_{LDA}(n) \qquad (30)$$

It is shown in Appendix B that a linear response calculation around the ground state performed on Eq. (26) using the definition of Eq. (29), is compatible with the GK linear response expression in Eq. (27).

The memory potential in Eq. (29) is still not TC. So we must correct it using the method of section II (15), giving the following memory functional:

$$v_{MEM-TC}[n](\mathbf{R},t) = \int_0^t F_{GK}(n(\mathbf{R},t),t-t') \\ \dot{n}(\mathbf{R} + \mathbf{D}[n](t') - \mathbf{D}[n](t),t')dt', \qquad (31)$$

We chose to demonstrate the memory effects on a cluster of Na$_{21}^+$ at zero temperature. The choice of this cluster size is its near spherical shape[15] and we have made sure the behavior depicted here is typical of other sodium clusters as well. Experimental absorption measurements of Na$_{21}^+$ were published by Schmidt et al[16], however these measurements are performed in temperatures were inhomogeneous effects are non-negligible and so we defer the comparison of our results with these experiment to a future, more detailed paper.

Two model cases of Na$_{21}^+$ are studied. One is a jellium sphere, with almost uniform ionic distribution,

$$n_+(\mathbf{R}) = \frac{n_0}{1 + e^{(R-r_c)/w}} \qquad (32)$$

Here we chose the parameter values $w = 0.4a_0$, $r_c = 10.8a_0$ and $n_0 = 0.004ea_0^{-3}$. These parameters ensure a "bulk" density corresponding to sodium and a total charge of $N_+ = 21e$. The second model is more realistic, where the positive charge is lumped as 21 atomic cores forming a cluster. In this case, a local pseudopotential (PP) is used[17] so that only valence electrons are considered. The DFT energy of the positive charge and 20 electrons is written as:

$$E[n] = T_s[n] + \int n(\mathbf{R})v_+(\mathbf{R})d^3R \\ + E_H[n] + E_{XC}[n] + E_{nuc} \qquad (33)$$

Here, $v_+(\mathbf{R})$ is the potential exerted by the positive charge on the electrons, $E_H[n] = \frac{1}{2}\int n(\mathbf{R})v_H[n](\mathbf{R})d^3R$ is the Hartree energy and $v_H[n](\mathbf{R}) = \int n(\mathbf{R}')/|\mathbf{R}-\mathbf{R}'|d^3R'$ is the Hartree potential. The approximation we use for the XC energy is $E_{LDA}[n] = \int n(\mathbf{R})\varepsilon_{hom}(n(\mathbf{R}))d^3R$ where $\varepsilon_{hom}(n)$ is the XC energy per particle of a homogeneous electron gas at density $n$, parameterized by the Perdew and Wang[18]. Finally, $E_{nuc}$ is the nuclear repulsion energy. In the case of the atomic cluster, both $v_+$ and $E_{nuc}$ are explicit functions of the atomic core positions. The placement of the atomic cores in the cluster is thus determined by the standard process of minimizing the total energy $E[n]$ (sum of electronic and nuclear repulsion). Specifically, we start from a well-known atomic configuration, reported in Ref.[15], and refined it using our code. This minimization process caused only slight rearrangement.

The absorption spectrum is determined by a procedure similar to the method depicted in Refs. [10, 19], where starting from the ground-state orbitals of the optimized structure, the KS orbitals are propagated in time according to the nonlinear time-dependent Schrodinger equation (the so-called time-dependent Kohn-Sham equations):

$$i\dot{\psi}_j(\mathbf{R},t) = -\frac{1}{2}\nabla^2\psi_j(\mathbf{R},t) + v_s[n](\mathbf{R},t)\psi_j(\mathbf{R},t); \\ j = 1,...,N_e/2 \qquad (34)$$

With the density given by the orbitals densities according to:

$$n(\mathbf{R},t) = 2\sum_{j=1}^{N_e/2}|\psi_j|^2 \qquad (35)$$

(closed shell is assumed for simplicity). The effective potential $v_s$ is written as follows:

$$v_s[n](\mathbf{R},t) = v_+(\mathbf{R}) + v_p(\mathbf{R},t) \\ + v_H[n(t)](\mathbf{R}) + v_{LDA}[n(t)](\mathbf{R}) \\ + v_{MEM}[n](\mathbf{R}) \qquad (36)$$

Here the driving TD perturbation is an impulsive (almost delta-function) dipole-coupled electric field pulse:

$$v_p(\mathbf{R},t) = -E_0 Z e^{-(t-t_0)^2/2\sigma^2} \qquad (37)$$

With parameters:

$$E_0 = 10^{-3} E_h a_0^{-1}, \quad t_0 = 8\hbar E_h^{-1}, \quad \sigma = 2a_0. \quad (38)$$

The memory potential $v_{MEM}$ is based on the Gross-Kohn (GK) XC kernel (see Eq. (29) and (31)), with different combinations of symmetry corrections, as described below.

**TABLE 1: THE NOTATION FOR COMBINATIONS OF THE GK MEMORY POTENTIAL CORRECTED FOR TC AND/OR THE ZERO-FORCE CONDITION**

|  | 0-force not enforced | 0-force enforced: Homogeneous field | 0-force enforced: Minimal tampering |
|---|---|---|---|
| **TC** | TC/0 | TC/1 | TC/2 |
| **Non-TC** | GK | GK/1 | GK/2 |

For each of the two $Na_{21}^+$ models, we compared the time-dependent dipole obtained from calculations using four different XC potentials. The nomenclature we use for the various potentials is shown in Table 1.

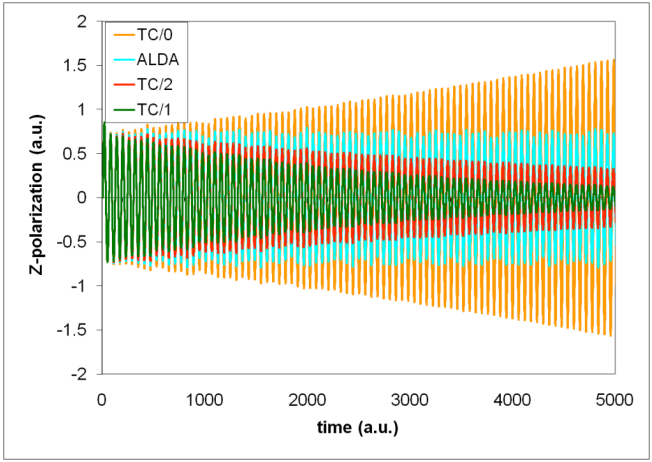

**FIGURE 1: TIME-DEPENDENT ELECTRONIC DIPOLE MOMENT (POLARIZATION IN Z DIRECTION) FOR JELLIUM MODEL OF THE $NA^+_{21}$ CLUSTER CALCULATED WITH TC POTENTIALS.**

We first study the dipole spectrum of $Na_{21}^+$ obtained from Gross-Kohn based potentials which are translationally covariant. Two models for the $Na_{21}^+$ cluster are considered: a spherical Jellium model, with results shown in Figure 1 Figure 1and a more realistic atomistic model with results shown in Figure 2Figure 2. For reference we include in each figure also the signal from adiabatic ALDA (that is with $v_{MEM}$ set to zero in Eq. (36)) which does not decay. We consider 4 types of corrected potentials. The signal computed by TC and the two zero-force potentials is stable and decaying. However, when one neglects to correct for zero force the resulting signal (TC/0) exhibits a non-physical growth. It is difficult to determine if this is an exponential growth resulting from instability or a strong but spurious low frequency mode. The results for a more realistic, atomic model of the $Na_{21}^+$ cluster, are shown in Figure 2. In this case, the spurious low-energy mode has luckily disappeared when the zero-force correction is not applied. It is interesting to note that the plasmon decay due to memory is considerably faster when a realistic atomistic potential is used. This is probably due to the fact, arising from the Harmonic Potential Theorem[4] (HPT), that there can be no memory effects in perfectly Harmonic external potentials. Inside spherical Jellium, the external potential is harmonic and so anharmonicity is weak, felt only near the surface, thus memory effects are not large.

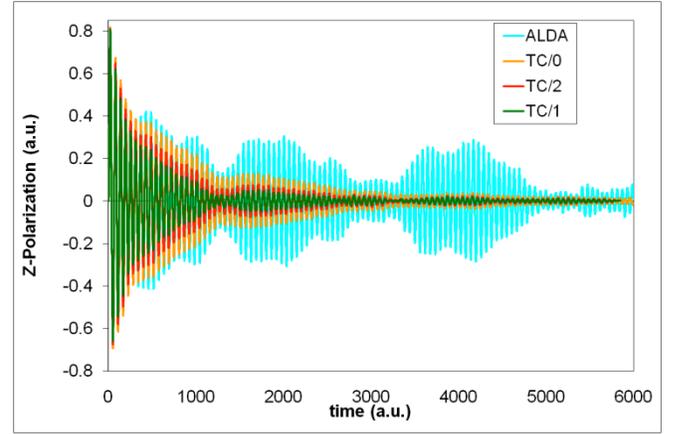

**FIGURE 2: TIME-DEPENDENT DIPOLE MOMENT FOR ATOMIC $NA_{21}^+$ CALCULATED WITH TC POTENTIALS.**

Consider now what can happen when TC is not observed. In Figure 3, using the Jellium model we see the hierarchy is exactly opposite to the previous case. Here, the strongest decay belongs to the uncorrected potential, $v_{GK}$ (red line). On the other hand if one corrects for the zero force condition, the results are very sensitive to the way one done this, as clearly seen when comparing the signal for GK/2 (decaying) and GK/1 (growing).

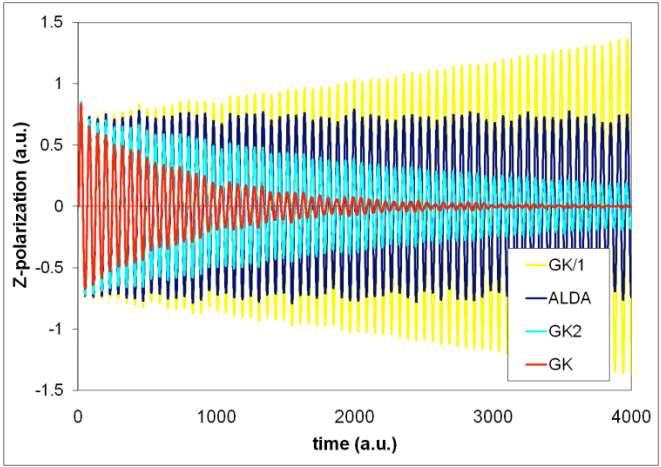

**FIGURE 3: TIME-DEPENDENT ELECTRONIC DIPOLE MOMENT FOR JELLIUM MODEL OF $NA_{21}^+$ CALCULATED WITH NON-TC POTENTIALS.**

The situations is once again very different when atomic clusters are used shown in Figure 4. Here the decay of all signals is faster, as before but the large sensitivity to the correction of the zero-force condition is smaller. It is clear that the TC correction is vital and only when TC is enforced do the results depend weakly on the way the 0-force is enforced.

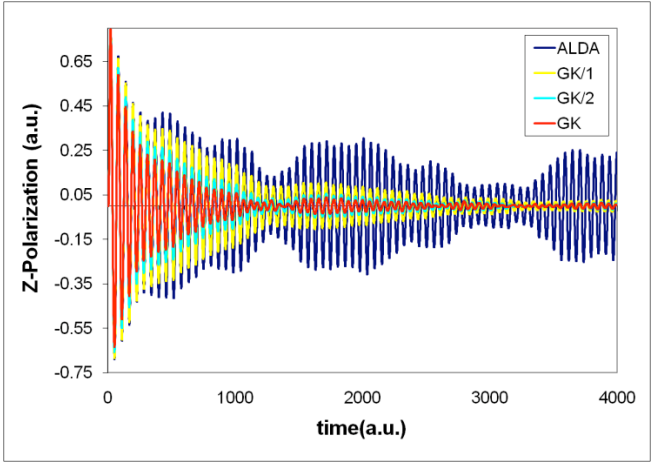

**FIGURE 4: TIME-DEPENDENT DIPOLE MOMENT FOR ATOMIC $NA_{21}^+$ CALCULATED WITH NON-TC POTENTIALS.**

## V. DISCUSSION AND CONCLUSIONS

We have presented the theory for imposing two basic symmetry constraints: translational covariance (TC) and zero force on a given memory functional. We also gave the expressions for linear response kernels in these cases. We discussed two ways to impose the zero force condition. One is through the imposing of a homogeneous electric field, resulting in a highly non-local potential and the second method involves imposition of a local potential depending on the gradients of the density.

We have presented calculations on two (Jellium and atomistic) models of the Na21+ cluster showing the effects memory has on the dipole response. Our results show that enforcing TC and 0-force conditions severely affects the dynamics of the system. These effects emerge especially in the case of the Jellium sphere, as is shown in Figure 1, where the neglect of enforcing the 0-force condition can result in developing spurious low frequency modes. If we preserve TC but not the zero force condition, an unstable mode can also arise (Figure 3). The emergence of a low mode can be explained by analytical considerations given in appendix C.

Another interesting point arises while comparing between the Na21+ Jellium and the atomic clusters. Consistently, for the corrected TC/1 and TC/2 potentials in Figure 1 the dipole decays much faster when the atomic cluster is corrugated by a local atomic potential. We have explained this striking difference using as resulting from the near-Harmonic potential existing in the Jellium sphere. Perfectly Harmonic potentials exhibit no decay[4].

The methods developed here can be applied to many suggestions existing in the literature of non-TC and non-zero force kernels[3, 14, 20] It can also be used in conjunctions with kernels recently developed by many-body perturbation theory approaches[21-23].

Acknowledgements: We gratefully acknowledge funding by a grant from the US-Israel Binational Science Foundation.

## APPENDIX A: TC-CORRECTION TO THE RESPONSE KERNEL

In this appendix we give the form of the correction required to turn a non-TC kernel into a TC kernel. We start with a general linear response case and then specialize to the linear response of the ground-state.

We want to know what is the XC kernel $f_{XC}(\mathbf{Y}_2;\mathbf{Y}_1) \equiv f_{XC}(\mathbf{R}_2,t_2;\mathbf{R}_1,t_1)$ describing the response of the XC potential at space-time point $\mathbf{Y}_2 \equiv (\mathbf{R}_2,t_2)$ to a small change in the density at space-time point $\mathbf{Y}_1 \equiv (\mathbf{R}_1,t_1)$:

$$\delta v_{XC}(\mathbf{Y}_2) = \int_0^{t_2}\int f_{XC}[n](\mathbf{Y}_2;\mathbf{Y}_1)\delta n(\mathbf{Y}_1)d^4Y_1 \quad (A.1)$$

We assume we are given a non-TC kernel $F_{XC}(\mathbf{y}_2,\mathbf{y}_1)$ corresponding to a non-TC potential $V_{XC}[n](\mathbf{y})$ (here we use the abbreviation $\mathbf{y} = (\mathbf{R}-\mathbf{D}(t),t)$). We use the fact that a non-TC potential can be corrected as depicted in Eq. (15). Taking the variation of this equation we have:

$$\begin{aligned}\delta v_{XC}[n](\mathbf{Y}) &= v_{XC}[n+\delta n](\mathbf{Y}) - V_{XC}[n](\mathbf{Y}) \\ &= V_{XC}[N+\delta N](\mathbf{y}) - V_{XC}[N](\mathbf{y}) \\ &\quad - \delta \mathbf{D} \cdot \nabla V_{XC}(\mathbf{y}) \end{aligned}$$
(A.2)

Working out the variations in detail, it is possible to show that the TC kernel $f_{XC}$ is constructed from the non-TC kernel $F_{XC}$ according to:

$$f_{XC}(\mathbf{Y};\mathbf{Y}') = F_{XC}(\mathbf{y},\mathbf{y}') + \frac{1}{N_e}\mathbf{R}' \cdot \mathbf{K}(\mathbf{y},t') \quad (A.3)$$

Where the correction term is determined by:

$$\mathbf{K}(\mathbf{y},t') = \int F_{XC}(\mathbf{r},t;\mathbf{r}'',t')\nabla N(\mathbf{r}'',t')dr'' \\ -\delta(t-t')\nabla V_{XC}[N](\mathbf{y})$$
(A.4)

We already know that adiabatic potentials are automatically TC. Let's see if the correction $\mathbf{K}$ is indeed zero in this case. When the potential functional is adiabatic we have $v_{XC}[n](\mathbf{R},t) = w(n(\mathbf{R},t)) = w(N(\mathbf{r},t))$, for some function $w(n)$. In this case, it is straightforward to show that:

$$F_{XC}(\mathbf{y},\mathbf{y}'') = \delta(t-t'')\delta(\mathbf{r}-\mathbf{r}'')w'(N(\mathbf{y})) \quad (A.5)$$

Plugging this in Eq. (A.4) gives after some manipulations $\mathbf{K} = 0$ in this adiabatic case, as required.

## TC-CORRECTING THE STATIONARY KERNEL

The correction term in Eq. (A.4) can be simplified if the response of the stationary ground state is considered as a special but important case. In this case, the kernels depend only on $t-t'$ and so we can set $t'=0$. Also, we the unperturbed CM is stationary so we can set $\mathbf{D}=0$ (making $\mathbf{r}$ and $\mathbf{R}$ identical). Thus from Eq. (15) we obtain:

$$\mathbf{K}(\mathbf{Y}) = \int F_{XC}^{stat}(\mathbf{R},\mathbf{R}'',t)\nabla n_{gs}(\mathbf{R}'')d\mathbf{R}'' \\ -\delta(t)\nabla V_{XC,gs}(\mathbf{R})$$
(A.6)

One can Fourier transform the time variable to a frequency and obtain:

$$\tilde{f}_{XC}(\mathbf{R},\mathbf{R}';\omega) = \tilde{F}_{XC}(\mathbf{R},\mathbf{R}';\omega) + \frac{1}{N_e}\mathbf{R}' \cdot \tilde{\mathbf{K}}(\mathbf{y},\omega)$$
(A.7)

Where the correction term is now simpler:

$$\tilde{\mathbf{K}}(\mathbf{Y},\omega) = \int \tilde{F}_{XC}^{stat}(\mathbf{R},\mathbf{R}'',\omega)\nabla n_{gs}(\mathbf{R}'')d\mathbf{R}'' \\ -\nabla V_{XC,gs}(\mathbf{R})$$
(A.8)

Note that the correction term is now dependent non-locally on position and on the gradient of the density. This is a direct and necessary result of memory.

## APPENDIX B: THE MEMORY KERNEL

In this appendix we study the relation between the XC kernel and the adiabatic part and non-adiabatic part of the XC potential, in detail.

We construct a memory including the XC potential functional $F(n,\tau)$ in the following way:

$$v_{XC}(\mathbf{R},t) = \int_0^t F(n(\mathbf{R},t),t-t')\dot{n}(\mathbf{R},t')dt' \\ + v_{AD}(n(\mathbf{R},t))$$
(B.1)

We now associate both the adiabatic part $v_{AD}$ and the kernel dependent part $F(n,\tau)$ with the homogeneous electron gas XC kernel $f(n,\tau) \equiv f_{XC}^h(n,\tau)$ defined through the linear response relation:

$$\delta v_{XC}(n,t) = \int_0^t f(n,t-t')\delta n(t')dt' \quad (B.2)$$

As is customary we use the Fourier transform of $f$:

$$\tilde{f}(n,\omega) = \int_0^\infty f(n,t)e^{i\omega t}dt \quad (B.3)$$

To get the relation between $f$ and $F$, we take the linear response of (B.1) around the ground state. We then have:

$$\delta v_{XC}(\mathbf{R},t) = \int_0^t F(n(\mathbf{R}),t-t')\delta\dot{n}(\mathbf{R},t')dt' \\ + v'_{AD}(\mathbf{R},t)\delta n(\mathbf{R},t).$$
(B.4)

Since everything is local, we drop henceforth the $\mathbf{R}$ notation. After integration by parts we obtain:

$$\delta v_{XC}(t) = \int_0^t \dot{F}(n, t-t')\delta n(t')dt' \quad \text{(B.5)}$$
$$+ [F(n,0) + v'_{AD}(n)]\delta n(t)$$

Comparing with Eq. (B.2) we find

$$f(n, t-t'') = \dot{F}(n, t-t'')\theta(t-t'')$$
$$+ [F(n,0) + v'_{AD}(n)]\delta(t-t'') \quad \text{(B.6)}$$

In order for all this to make sense we must demand 2 things:

$$\dot{F}(n,\tau) = f(n,\tau) \quad \text{(B.7)}$$

And:

$$F(n,0) = -v_{AD}(n) \quad \text{(B.8)}$$

Another constraint is that we want to have a finite memory, i.e.:

$$\lim_{\tau \to \infty} F(n,\tau) = 0 \quad \text{(B.9)}$$

From Eq. (B.7) we write also:

$$F(n,\tau) = F(n,0) + \int_0^\tau f(n,\tau')d\tau' \quad \tau \geq 0 \quad \text{(B.10)}$$

Inserting $\tau \to \infty$ and using Eq. (B.9), we find:

$$F(n,0) = -\int_0^\infty f(n,t')dt' \quad \text{(B.11)}$$

The integral on the right is simply the DC term n the frequency kernel $\tilde{f}(n,0)$. Thus. a stringent constraint on the compatibility of $v_{AD}$ with $f$ is:

$$v'_{AD}(n) = \tilde{f}(n,0) \quad \text{(B.12)}$$

We note that $GK$ kernel acknowledges this and constructs $\tilde{f}_{GK}$ to obey Eq. (B.12) for the LDA potential. To summarize, $f_{XC}$ determines both the adiabatic potential $v_{AD}$ through (B.12) and the memory part of the XC potential through:

$$F(n,\tau) = \theta(\tau)\left[\int_0^\tau f_{GK}(n,\tau')d\tau' - v'_{AD}(n)\right] \quad \text{(B.13)}$$

Since all the arguments are valid for the homogeneous electron gas, we must have in this development $v_{AD}(n) = v_{LDA}(n)$. As noted above, the Gross-Kohn XC kernel indeed satisfies this constraint[3].

## APPENDIX C: SPURIOUS MODES

In this appendix we show why spurious modes may arise when 0-force is not enforced. Consider a Harmonic potential $v_{harm}(\mathbf{R}) = \frac{1}{2}\mathbf{R}^T \overset{\leftrightarrow}{\mathcal{H}} \mathbf{R}$ and a homogeneous time-dependent external field $\mathbf{E}(t)$. Now apply Ehrenfest's theorem to the electronic center of mass $\mathbf{D}(t)$ (Eq. (10)) in the time-dependent Kohn-Sham equations (Eqs. (34)). The Hartree and LDA potentials do not exert a net force so the equation of motion for $\mathbf{D}$ is:

$$\ddot{\mathbf{D}}(t) = -\overset{\leftrightarrow}{\mathcal{H}}\mathbf{D}(t) + \mathbf{E}(t)$$
$$- \int n(\mathbf{R},t)\vec{\nabla}v_{Mem}(\mathbf{R},t)d^3R \quad \text{(C.1)}$$

Here we assume that the memory $v_{Mel}$ is TC but does not obey the 0-force condition. Denote $n_0(\mathbf{R})$ the ground-state density and linearized to the first order density response $n_1(\mathbf{r},t)$. The dipole moment is a 1$^{st}$ order quantity, $\mathbf{D}_1 = \frac{1}{N_e}\int n_1(\mathbf{R},t)\mathbf{R}d^3R$. Then, expanding Eq. (29) to first order:

$$v_{Mem}(\mathbf{R},t) = \theta(t)\int_0^t F_{XC}(n_0(\mathbf{R}), t-t') \times$$
$$[\dot{n}_1(\mathbf{R},t') + \dot{\mathbf{D}}_1(t') \cdot \nabla n_0(\mathbf{R})]dt' \quad \text{(C.2)}$$

We find the memory potential is itself a 1$^{st}$ order quantity. Therefore, the half Fourier transformation of Eq. (C.1), to 1$^{st}$ order is:

$$\omega^2 \tilde{\mathbf{D}}_1 = \overset{\leftrightarrow}{h}\tilde{\mathbf{D}}_1 - \tilde{\mathbf{E}} + \int n_0(\mathbf{R})\nabla \tilde{v}_{Mem}(\mathbf{R},\omega)d^3R \quad \text{(C.3)}$$

From (C.2) the half Fourier transform of $v_M$ is:

$$\tilde{v}_{Mem}(\mathbf{R},\omega) = i\omega \tilde{F}_{XC}(n_0,\omega)[\tilde{n}_1(\omega) + \tilde{\mathbf{D}}_1(\omega) \cdot \nabla n_0] \quad \text{(C.4)}$$

Substituting the potential in Eq. (C.4) into Eq. (C.3), we obtain the following form:

$$\tilde{\mathbf{D}}_1 = \frac{\tilde{\mathbf{Q}}}{\left(\vec{\vec{\mathcal{H}}} + i\omega\vec{\vec{\Omega}} - \omega^2\right)} \tag{C.5}$$

Where:

$$\begin{aligned}\vec{\vec{\Omega}} &\equiv \int \tilde{F}_{XC}(n_0,\omega)\left(\vec{\nabla}n_0\vec{\nabla}n_0\right)d^3R \\ \tilde{\mathbf{Q}} &\equiv \tilde{\mathbf{E}} + i\omega \int \tilde{F}_{XC}(n_0,\omega)\vec{\nabla}n_0(\mathbf{R})\tilde{n}_1(\omega)d^3R\end{aligned} \tag{C.6}$$

Both the tensor $\vec{\vec{\Omega}}$ and the shift $\tilde{\mathbf{Q}}$ are zero if the memory potential obeys the zero-force condition. In this case the only poles in $\mathbf{D}_1(\omega)$ are those due to the eigenvalues of the Harmonic hessian $\vec{\vec{\mathcal{H}}}$. This is the content of the Harmonic potential theorem in linear response. However when the zero-force condition is not obeyed we immediately see that the poles may either shift or even obtain decaying (positive imaginary poles) or exploding characteristics (negative imaginary poles), depending on the tensor $\vec{\vec{\Omega}}$.